\documentclass[12pt,a4paper]{article}
\usepackage[utf8]{inputenc}
\usepackage[T1]{fontenc}
\usepackage{amsmath}
\usepackage{amsfonts}
\usepackage{amssymb,amsthm}
\usepackage{graphicx}
\usepackage{mathrsfs}
\usepackage{hyperref}
\usepackage{color}
\usepackage{bigints}
\usepackage{etoolbox}
\usepackage{tikz}
\usepackage{subcaption} 
\usepackage{comment}

\newtheorem{theorem}{Theorem}

\newtheorem{example}[theorem]{Example}

\newtheorem{proposition}[theorem]{Proposition}
\newtheorem{remark}[theorem]{Remark}

\newcommand{\R}{\mathbb{R}}
\newcommand{\N}{\mathbb{N}}

\def\A{\mathcal{A}}
\def\div{\operatorname{div}}
\def\grad{\operatorname{grad}}
\def\curl{\operatorname{curl}}
\def\Div{\operatorname{Div}}
\def\dist{\operatorname{dist}}

\def\N{\mathcal{N}}

\def\Ker{\operatorname{Ker}}
\def\H{\mathbb{H}}

\def\R{\mathbb{R}}
\def\SI{\operatorname{SI}}
\def\Sc{\operatorname{Sc}}
\def\Vec{\operatorname{Vec}}
\def\Sol{\operatorname{Sol}}
\def\Irr{\operatorname{Irr}}
\def\tr{\operatorname{\rm tr}}

\def\nrm{_{\mathbf n}}

\def\Ti{T_{0,\Omega}}
\def\Tii{{\overrightarrow T}_{\!\!1,\Omega}}
\def\Tiii{\overrightarrow T_{\!\!2,\Omega}} 
\def \U {\overrightarrow U_{\!\!\Omega} }

\makeatletter
\patchcmd{\maketitle}{\@fnsymbol}{\@alph}{}{}  
\makeatother
\title{A right-inverse of curl which is divergence-free invariant and some applications to generalized Vekua type problems}

\makeatletter
\newcounter{author}
\renewcommand*\author[1]{%
\stepcounter{author}%
\ifnum\c@author=1
\gdef\@author{#1}%
\else
\xdef\@author{\unexpanded\expandafter{\@author\and#1}}%
\fi
\csgdef{author@\the\c@author}{#1}}
\newcommand*\email[1]{%
\csgdef{email@\the\c@author}{#1}}
\newcommand*\address[1]{%
\csgdef{address@\the\c@author}{#1}}
\AtEndDocument{%
\xdef\author@count{\the\c@author}%
\c@author=1
\print@authors}
\newcommand*\print@authors{%
\ifnum\c@author>\author@count
\else
\print@author{\the\c@author}%
\advance\c@author by 1
\expandafter\print@authors
\fi}
\newcommand*\print@author[1]{%
\par\medskip
\begin{tabular}{@{}l@{}}%
\textsc{Addresses of \csuse{author@#1}}\\
\csuse{address@#1}\\
\textit{E-mail address}:
\href{mailto:\csuse{email@#1}}{\csuse{email@#1}}
\end{tabular}}
\makeatother
{\small
\author{Briceyda B. Delgado}
\address{Universidad Autónoma de Aguascalientes, Aguascalientes, Mexico}
\email{briceydadelgado@gmail.com}

\author{J. E. Mac\'{\i}as-D\'{\i}az}
\address{Universidad Aut\'{o}noma de Aguascalientes, Aguascalientes, Mexico\\Tallinn University, Tallinn, Estonia}
\email{jemacias@correo.uaa.mx}

}

\begin{document}

\maketitle
{ }\makeatletter{\renewcommand*{\@makefnmark}{}
} 

{ }\makeatletter{\renewcommand*{\@makefnmark}{}
}

\begin{abstract}
  In this work, we investigate the system formed by the equations $\div \vec w=g_0$ and $\curl \vec w=\vec g$ in bounded star-shaped domains of $\R^3$. A Helmholtz-type decomposition theorem is established based on a general solution of the above-mentioned div-curl system which was previously derived in the literature. When $g_0\equiv 0$, we readily obtain a bounded right inverse of $\curl$ which is a divergence-free invariant. The restriction of this operator to the subspace of divergence-free vector fields with vanishing normal trace is the well-known Biot--Savart operator. In turn, this right inverse of $\curl$ will be modified to guarantee its compactness and satisfy suitable boundary-value problems. Applications to Beltrami fields, Vekua-type problems as well as Maxwell's equations in inhomogeneous media are included.
\end{abstract}


\section{Introduction\label{sec:introduction}}

One of the fundamental theorems in vector analysis is the very well-known ``Helmholtz Decomposition Theorem''. This result states that any vector field in $\R^3$ can be completely characterized in terms of its divergence and curl (sometimes also called rotational or vorticity). 
This theorem was formulated by Hermann von Helmholtz \cite{Helmholtz1873}, and it represents any vector field $\vec w$ in $\R^3$ as the sum of a divergence-free vector field $(\curl \vec v)$ and an irrotational vector field $(\grad v_0)$. More precisely, the following decomposition is satisfied:
\begin{equation}
  \label{eq:Helmholtz-R3}%
  \vec w = - \grad v _0 + \curl \vec v.
\end{equation} 
Here, the \textit{Helmholtz potentials} $v_0$ and $\vec v$ are given (see \cite[p. 166]{Korn1968} and \cite[Thm. 5.1.1]{GuHaSpr2016}) by  
\begin{align}
  v _0(\vec x) & = - L _{\R ^3} [\div \vec w] (\vec x), \\
	\vec v (\vec x) & = - L _{\R ^3} [\curl \vec w] (\vec x).
\end{align}
In turn, $L$ is the \textit{Newton potential} is defined by
\begin{align}
   L _{\R ^3} [u] (\vec {x}) = - \frac {1} {4 \pi} \int _{\R ^3}\frac{{u (\vec{y})}} {\left\vert {\vec {x} - \vec {y}} \right\vert}{d \vec {y}},
\end{align}	
and it is a right inverse of the Laplacian. Later on, the uniqueness of the Helmholtz decomposition \eqref {eq:Helmholtz-R3} was proved under the assumption that the solution satisfies the asymptotic behavior $w(\vec x)=o(1/|\vec x|)$, for $|\vec x|\rightarrow \infty$. In particular, it was established that
\begin{equation}
  \label{eq:right-inverse-R3}
  \curl \vec v(\vec x)=\curl \frac{1}{4\pi}\int_{\R^3}\frac{{\curl \vec w(\vec{y})}}{\left\vert {\vec{x}-\vec{y}}\right\vert }{\,d\vec{y}}
  =-\int_{\R^3} \dfrac{(\vec x-\vec y)\times \curl \vec w(\vec y)}{4\pi |\vec x-\vec y|^3}\, d\vec y,  
\end{equation}
is a right inverse of $\curl$ in the entire three-dimensional space. It is worth pointing out here that \eqref{eq:right-inverse-R3} is sometimes called the \textit{Biot--Savart operator}.

In this work, we will use techniques employed in various works concerning the div-curl system in star-shaped domains \cite {DelPor2017}, the div-curl system in Lipschitz domains \cite {DelPor2018}, a perturbed div-curl system \cite {DelKrav2019} and the div-curl system in exterior domains \cite {DelMac2021}. In all of those works, the quaternionic analysis played a fundamental role. In particular, we will show that any vector field $\vec w$ admits a Helmholtz-type decomposition in bounded star-shaped domains (see Proposition \ref{prop:Helmholtz-deco} below) as follows:
\begin{equation}
  \label{eq:deco-Helmholtz-intro}
  \vec w=\grad v_0^*-\curl \vec v^*.
\end{equation} 
Here, the potentials $v_0^*$ and $\vec v^*$ are defined over $\Omega$, and are given by
\begin{equation}
  \label{eq:new-potentials}
  \left\{ \begin{aligned}
    v_0^*(x)&=\int_{\Omega} \dfrac{\div \vec w(\vec y)}{4\pi|\vec x-\vec y|}\, dy,\\	
    \vec v^*(\vec x)&=\int_{\Omega}\dfrac{\curl \vec w(\vec y)}{4\pi|\vec x-\vec y|}\, dy	+\int_0^1 \frac{t|\vec x|^2}{2} \grad \psi_0(t\vec x) \, dt.
  \end{aligned} \right.
\end{equation}
Moreover,
\begin{equation}
  \label{eq:psi-0}
  \psi_0(\vec x)=-\div_{\vec x} \int_{\Omega} \frac{\curl \vec w(\vec y)}{4\pi|\vec x-\vec y|}\, d\vec y. 
\end{equation}	

In addition, $\div \vec v^*$ is harmonic in $\Omega$. Some examples of this solution are computed implicitly in Example \ref{ex:constant}. Comparing \eqref{eq:Helmholtz-R3} and \eqref{eq:deco-Helmholtz-intro}, we can see that the potentials are now given as integrals over the domain $\Omega$ instead of over the entire three-dimensional space. Besides, our potential $\vec v^*$ has an extra term, we will analyze in detail the operator which represents it, as it is the key for the analysis of some boundary-value problems associated to the div-curl system.
	
Analogizing the operators involved in the Helmholtz-type decomposition \eqref{eq:deco-Helmholtz-intro}--\eqref{eq:psi-0} with one of the most important operators in quaternionic analysis ---the Teodorescu transform $T_{\Omega}$---, the following is an appropriate decomposition:
\begin{equation}
  T_{\Omega}[-\div \vec w+\curl \vec w]=\Ti[\curl \vec w]-\Tii[\div \vec w]+\Tiii[\curl w],
\end{equation}
where
\begin{align}
  \Ti[\curl  \vec w] &=\psi_0, \\
  \Tii[\div w] & =\grad v_0^*, \\
  \Tiii[\curl \vec w] & =\curl \vec v^*-\curl \int_0^1 \frac{t|\vec x|^2}{2} \grad \psi_0(t\vec x) \, dt,
\end{align}	
and $v_0^*$, $\vec v^*$ and $\psi_0$ are as previously defined in \eqref{eq:new-potentials} and \eqref{eq:psi-0}, respectively.
	
One of the aims of this work is to study analytically the solutions of the div-curl system in star-shaped domains under appropriate boundary conditions. In that sense, this manuscript may be considered as a completion of the analysis carried out in \cite{DelPor2017}. It is important to point out here that a general solution of the div-curl system in star-shaped domains was obtained in \cite{DelPor2017} assuming no boundary conditions. The explicit solution of that problem is recalled below in Theorem \ref{th:div-curl}, and it is given in terms of some integral operators appearing in quaternionic analysis as well as a monogenic completion operator. The solution of the problem under consideration in the present work hinges on embedding the vector-valued differential system into a quaternionic-valued one. A solution is found in the algebra $\H$ of quaternions, and a projection into $\R^3$ is performed then without affecting the equivalent system. As a consequence of the Helmholtz-type decomposition  \eqref{eq:deco-Helmholtz-intro}-\eqref{eq:psi-0}, we obtain a right inverse of $\curl$. The fact that this right inverse is not unique as well as its regularity properties will be used to satisfy some suitable boundary-value problems (BVPs). These results will be presented in Section~\ref{sec:BVP}. Throughout, we will work in the context of the spaces $W^{p,\mbox{\scriptsize div-curl}}(\Omega)$ and $W_{\nrm}^{p,\mbox{\scriptsize div-curl}}(\Omega)$ in order to guarantee a natural regularity for all $L^p$-solutions.

It is important to mention that there are several results in the literature on the existence of an inverse of $\curl$. To that end, authors usually impose some suitable additional boundary conditions in bounded domains\cite{Girault1986, Yoshida1990}. In particular, the existence of a compact inverse operator of $\curl$ in a subspace of the divergence-free vector fields with vanishing normal trace was proved in \cite{Yoshida1990}. The boundary conditions used in that work assured that $\curl$ and $\curl^{-1}$ were self-adjoint operators in their respective domains of definition. In the present work, we will provide an explicit expression of a right inverse of $\curl$ from the subspace of divergence-free vector fields to the subspace of divergence-free vector fields with vanishing normal trace. The right inverse preserves the property of being compact (see Proposition \ref{th:right-inverse-compact} below). It is worth mentioning that the domains of definition of $\curl$ and its right inverse will be larger than that considered in \cite{Yoshida1990}. 

On the other hand, the BVP 
\begin{equation}
  \left\{ \begin{aligned}
    \curl \vec w & =\vec g, \\
    \vec w|_{\partial\Omega} & = 0,
  \end{aligned} \right.
\end{equation}
has been extensively studied, see for instance \cite {Borchers1990, Griesinger1990, Wahl1990}. In particular, the author of \cite {Griesinger1990} provided necessary and sufficient conditions for the existence of a unique solution which depends continuously on $\vec g$ in bounded domains of $\R ^n$. More recently, the authors of \cite{Berselli2019} found a solution in star-shaped domains with respect to a ball, under the assumption that $\vec g$ is a divergence-free Dini-continuous function. That solution was expressed in terms of an integral representation formula which, in turn, was obtained in \cite {Griesinger1990}. In our construction of a right inverse of $\curl$ that vanishes on the boundary, the monogenic completion process involved in the construction of the right inverse of $\curl$ without boundary conditions is straightforward because the scalar part of the Teodorescu transform applied to $\vec g$ vanishes. In the alternative proof that we present herein, we will follow some ideas used in \cite{Borchers1990}. As an application of our right inverse of the $\curl$ operator (namely $R_{\Omega}$, which is bounded in $L^p$ and divergence-free invariant), we will construct \textit{Beltrami fields} with coefficient $\alpha_0$ through a Neumann series that converges uniformly for certain values of $\alpha_0$ satisfying $|\alpha_0|<1/\|R_{\Omega}\|$ (see Theorem \ref{th:construction-Beltrami-constant}).

The outline of this manuscript is as follows. In Section \ref{sec:div-curl}, we will present some preliminaries needed to construct the general solution of the div-curl system.
In Section \ref{sec:right-inverse-curl}, we will provide a Helmholtz-type decomposition in bounded star-shaped domains of $\R^3$, which generalizes the classical Helmholtz decomposition \eqref{eq:Helmholtz-R3}. Moreover, a right inverse of $\curl$ without boundary conditions is derived and we analyze some properties inherited from the Teodorescu transform, such as boundedness and compactness. In Subsection \ref{sec:BVP}, we will investigate BVPs for the equation $\curl \vec w=\vec g$, considering the Neumann boundary condition $\vec w|_{\partial\Omega}\cdot \eta=0$ or, alternatively, the Dirichlet boundary condition $\vec w|_{\partial\Omega}=0$. Section \ref{sec:Beltrami-fields} provides an application of the bounded right inverse of $\curl$ to the construction of Beltrami fields. More precisely, we analyze the eigenvalue problem $\curl \vec w = \alpha _0 \vec w$ without boundary conditions, and then with the boundary condition $\vec w | _{\partial \Omega}\cdot \eta = a _0$ as well (see Propositions \ref{th:construction-Beltrami-constant} and \ref{prop:BVP-Beltrami}, respectively). The last section of this work will be devoted to applying our results to several related problems, including some generalized Vekua-type problems and the Maxwell system in inhomogeneous media.



\section{Preliminaries\label{sec:div-curl}} 

\subsection{Function spaces}

In this work, we will employ the usual function spaces associated to the operators $\div$ and $\curl$, which appear in many problems on electromagnetism \cite{Dautray1985, Girault1986}. More precisely, we will let
\begin{align}
   W ^{p , \div} (\Omega , \R ^3) & = \{ \vec u \in L ^p (\Omega , \R ^3) : \div \vec u \in L ^p (\Omega , \R)\}, \\
   W ^{p , \curl} (\Omega , \R ^3) & = \{ \vec u \in L ^p (\Omega , \R ^3) : \curl \vec u \in L ^p (\Omega , \R ^3)\}.
\end{align}
It is well known that $W ^{p , \div} (\Omega , \R ^3)$ and $W ^{p , \curl} (\Omega , \R ^3)$ are Banach spaces with respect to the norms 
\begin{align}
   \|\vec u\|_{W^{p,\div}} & = \|\vec u\|_{L^p}+\|\div \vec u\|_{L^p}, \\
   \|\vec u\|_{W^{p,\curl}} & = \|\vec u\|_{L^p}+\|\curl \vec u\|_{L^p}.
\end{align}
The classical Sobolev space $W ^{1 , p} (\Omega , \R^3)$ is a proper subset of the intersection
\begin{align}
  W^{p,\,\mbox{\scriptsize  div-curl}}(\Omega,\R^3) = W^{p,\div}(\Omega,\R^3)\cap W^{p,\curl}(\Omega,\R^3).
\end{align}

The \textit{normal trace operator} in $W^{p,\div}(\Omega,\R^3)$ is the function $\gamma\nrm \colon W^{p,\div}(\Omega,\R^3) \to (W^{1-1/q,q} (\partial\Omega,\R))^*$ defined by $\gamma\nrm(\vec u) =\vec u|_{\partial\Omega} \cdot \eta$. By the Divergence Theorem, the normal trace operator is weakly defined by 
\begin{align}
  \label{eq:normal-trace}
  \langle \gamma\nrm(\vec u), \tr v_0 \rangle_{\partial\Omega}= \int_{\Omega}{\vec u \cdot \nabla v_0 \, d\vec y} +  \int_{\Omega}{(\div \vec u) v_0 \, d\vec y},
\end{align}
for each $\vec u \in W^{p,\div}(\Omega,\R^3)$ and each $v_0\in W^{1,q}(\Omega,\R)$. Here, the symbol $\langle \cdot, \cdot \rangle_{\partial\Omega}$ denotes the duality pairing between $W^{1-1/q,q}(\partial\Omega)$ and its dual space $(W^{1-1/q,q}(\partial\Omega))^*$. When $p=2$, we will denote $W^{1/2,2}(\partial\Omega)$ and $(W^{1/2,2}(\partial\Omega))^*$ by $H^{1/2}(\partial\Omega)$ and $H^{-1/2}(\partial\Omega)$, respectively.
Let $W_0^{p,\div}(\Omega,\R^3)$ be the kernel of the normal trace operator $\gamma\nrm$. That is, let
\begin{equation}
  W_0^{2,\div}(\Omega,\R^3)=\{\vec u\in W^{p,\div}(\Omega,\R^3)\colon \gamma\nrm(\vec u)=0\}.
\end{equation}
Define also the space
\begin{equation}
  \label{eq:kernels}
  W\nrm^{p,\div\text{-}\curl}(\Omega,\R^3)=W_0^{p,\div}(\Omega,\R^3)\cap W^{p,\curl}(\Omega,\R^3),
\end{equation}
and endow it with the norm $\|\vec u\|_{L^p}+\|\div \vec u\|_{L^p}+\|\curl \vec u\|_{L^p}$.

\subsection{Construction of solution}

In this stage of our work, we will recall some results reported in \cite{DelPor2017, Thesis2018}. In particular, we will employ the constructive solution obtained in those works, for the div-curl system in bounded star-shaped domains $\Omega\subseteq\R^3$. Consider the div-curl system without any boundary conditions
\begin{equation}
  \label{eq:div_curl_system}
  \begin{aligned}
    \div\vec w & = g_0, \quad \text{in } \Omega, \\ 
    \curl\vec w & = \vec g, \quad \text{in } \Omega,
  \end{aligned}
\end{equation}
where $g_0\in L^p(\Omega,\R)$, $\vec g\in L^p(\Omega, \R^3)$ and $\div \vec g=0$. Here, $\vec g$ is required to be \emph {weakly solenoidal}, that is,
\begin{equation}
    \int_{\Omega}{\vec g \cdot \nabla u_0 \, d\vec x}=0,
\end{equation}
for all test functions $u_0\in W_0^{1,q}(\Omega,\R)$ and $\frac {1} {p} + \frac {1} {q}=1$. 

Recall that the \textit{Moisil--Teodorescu} differential operator is defined by
\begin{equation}
  \label{eq:Moisil-Teodorescu}
  D=e_1\frac{\partial}{\partial x_1}+e_2\frac{\partial} {\partial x_2}+e_3\frac{\partial}{\partial x_3}.
\end{equation}
We say that $w\in C^1(\Omega,\H)$ is \textit{left-monogenic} (respectively, \textit{right-monogenic}) in $\Omega$ when $D w = 0$ (respectively, $w D = 0$). In the following, we convey that the term `monogenic function' will refer to left-monogenic functions. It is well-known that $\Delta=-D^2$. As a consequence, if $w$ is (left- or right-) monogenic then every component function $w_i$ is harmonic, for $i = 0$, $1$, $2$, $3$. The application of $D$ to differentiable functions of the form $w=w_0+\vec w$ yields $Dw=-\div \vec w+\grad w_0+\curl \vec w$. One of the fundamental features to derive the solution of \eqref{eq:div_curl_system} in \cite{DelPor2017} was that \eqref{eq:div_curl_system} can be rewritten in terms of the Moisil--Teodorescu operator as a quaternionic formula, namely,
\begin{equation}
   D \vec w = - g _0 + \vec g.
\end{equation}
Likewise, the use of quaternionic integral operators as well as a monogenic completion procedure were crucial in deriving the solution reported in \cite{DelPor2017}. Given a scalar harmonic function $u _0$, this completion process consisted in finding a purely vector harmonic function $\vec u$, such that $D (u _0 + \vec u) = 0$.

The \textit{Teodorescu transform} and the \textit{Cauchy operator} are defined respectively by
\begin{align}
  \label{eq:operador_Teodorescu}
   T_{\Omega}[w](\vec x) &= -\int_{\Omega} 
	  E(\vec y-\vec x) w(\vec y)\,d\vec y, \quad \vec x\in\R^3,\\
  \label{eq:operador_Cauchy}
   F_{\partial\Omega}[\varphi](\vec x)&=\int_{\partial\Omega} E(\vec y-\vec x) \eta(\vec y) \varphi(\vec y) \,ds_{\vec y}, \quad    \vec x\in\R^3\setminus \partial\Omega,
\end{align}
where $E(\vec x)=-\vec x/(4\pi|\vec x|^3)$, $\vec x\in \R^3\setminus\{0\}$ is the Cauchy kernel and $\vec \eta$ is the outward normal vector to $\partial\Omega$. We will usually work with $w\in L^p(\Omega)$ and $\varphi \in L^p(\partial\Omega)$. Following the notation of the decomposition used in \cite{DelPor2017}, we denote the component operators of the Teodorescu transform as follows:
\begin{equation}\label{eq:decomposition_T}
   T_{\Omega}[w_0+\vec w] = \Ti[\vec w] + 
	 \Tii[w_0]+\Tiii[\vec w].
\end{equation}
Here, the scalar part is given by
\begin{equation}\label{eq:T0}
  \Ti[\vec w](\vec x) = \int_{\Omega} E(\vec y-\vec x) \cdot \vec w(\vec y) \,d\vec y, 
\end{equation}
and its vector part is divided for strategic reasons as
\begin{align}
  \Tii[w_0](\vec x) & = -\int_{\Omega} w_0(\vec y) E(\vec y-\vec x)   \,d\vec y,  \\
  \Tiii[\vec w](\vec x) & = -\int_{\Omega} E(\vec y-\vec x) \times \vec  w(\vec y) \,d\vec y.
\end{align}

The \textit{monogenic completion operator} $\U \colon\text{Har}(\Omega,\R)\to\text{Har}(\Omega,\R^3)$ is given by
\begin{align}\label{def:operator_completation_monogenic}
   \U [w_0](\vec x) = \Vec\left(\int_0^1 t\vec x  D w_0(t\vec x)  \,dt\right) =
   \int_0^1 t \vec x \times  \nabla w_0(t\vec x)   \,dt, \quad \vec x \in \Omega.
\end{align} 
This function sends real-valued harmonic functions into vector-valued harmonic functions. It is defined on star-shaped open sets $\Omega$ with respect to the origin. When $\Omega$ is star-shaped with respect to some other point, the definition of $\U $ is adjusted by shifting the
values of $\vec x $ accordingly. It has been established that $w_0+\U [w_0]$ is monogenic, for any real-valued harmonic function $w_0$ (see \cite[Prop.\ 2.3]{DelPor2017}).

The following result was proved in \cite[Thm.\ 4.4]{DelPor2017}.

\begin{theorem}[Delgado and Porter \cite{DelPor2017}]
  \label{th:div-curl}%
  Let $\Omega$ be a star-shaped open set. Let $g=g_0+\vec g\in L^p(\Omega)$ and $\div \vec g=0$. Then a general weak solution of the div-curl system \eqref{eq:div_curl_system} is given by
  \begin{align}\label{eq:general_solution}
    \vec w = -\Tii[g_0] + \Tiii[\vec g] -\U [\Ti[\vec g]] + \nabla h,
  \end{align}
  where $h\in\text{Har}(\Omega,\R)$ is arbitrary.
\end{theorem}

\begin{remark}
  \label{rmk:T0-harmonic}%
  Notice that the operator $\U$ in \eqref{eq:general_solution} acts over $\Ti[\vec g]$, which is harmonic. Indeed, observe that
  \begin{equation}
    \Delta T_{\Omega}[\vec g]=-D^2 T_{\Omega}[\vec g]=-D\vec g=\div \vec g-\curl \vec g.
  \end{equation}
  Taking its scalar part, we check that $\Ti[\vec g]$ is harmonic if and only if $\div \vec g=0$. This fact was crucial in the construction of this general solution. An alternative construction was given in \cite[Appendix]{DelPor2018}, removing the restriction of star-shapedness. In fact, this alternative general solution is valid for bounded Lipschitz domains in $\R^3$ with weaker topological constraints. The difficulty in this case lies in the inversion of a layer potential that appears in its expression \cite[Th.\ A.1]{DelPor2018}. It is worth mentioning that this problem is not present in the derivation of \eqref{eq:general_solution}.
\end{remark}

\section{Methods}
\subsection{Helmholtz decomposition}

The div-curl system \eqref{eq:div_curl_system} has been studied from various points of view in light of its fundamental importance in physics. Unfortunately, the solution \eqref{eq:Helmholtz-R3} is provided by integral operators which are defined over all the three-dimensional space. This feature of that solution represents a serious limitation for many applications. In the following, we will obtain a Helmholtz-type decomposition for the solution given in Theorem \ref{th:div-curl}. As observed in \cite[Prop.\ 3.2]{DelPor2017}, the components of the Teodorescu operator $T_{\Omega}$ can be rewritten in terms of the Newton potential $L$ over $\Omega$ as
\begin{equation}\label{eq:deco-T}
  \begin{aligned}
    \Ti[\vec g] & =\div L[\vec g], \\
    \Tii[g_0] & =-\grad L[g_0], \\ 
	  \Tiii[\vec g] & =-\curl L[\vec g],
	\end{aligned}
\end{equation}
where 
\begin{align}
  \label{eq:Newton-potential}
  L[u](\vec{x})=-\frac{1}{4\pi}\int_{\Omega}\frac{{u(\vec{y})}} {\left\vert {\vec{x}-\vec{y}}\right\vert }{\,d\vec{y}}.
\end{align}	
Analogously, a similar decomposition was pointed out for the operator $\U$ (see \cite[Cor.\ 4.3]{DelPor2017}), namely,
\begin{equation}
  \label{eq:deco-U}
  \U [u_0]=\curl \int_0^1 \frac{|t\vec x|^2}{2t} \grad u_0(t\vec x)\, dt=\curl \int_0^1 \frac{t|\vec x|^2}{2} \grad u_0(t\vec x)\, dt.   
\end{equation}
The solution of the div-curl system can be rewritten in a way similar to the classic \textit{Helmholtz decomposition theorem}: 

\begin{proposition}\label{prop:Helmholtz-deco}
  Let $\Omega$ be a star-shaped open set, and let $g=g_0+\vec g\in L^p(\Omega)$ with $\div \vec g=0$. Then the solution \eqref{eq:general_solution} admits a Helmholtz-type decomposition
  \begin{equation}
    \label{eq:deco-Helmholtz}
    \vec w=\grad L[g_0]-\curl \vec v^*,
  \end{equation} 
  where $\vec v^*$ is given by
  \begin{equation}
    \vec v^*=L[\vec g]+\int_0^1 \frac{t|\vec x|^2}{2} \grad \Ti[\vec g](t\vec x) \, dt.
  \end{equation}
  Moreover, $\div \vec v^*$ is harmonic in $\Omega$.
\end{proposition}

\begin{proof}
  Replacing \eqref{eq:deco-T}--\eqref{eq:deco-U} in \eqref{eq:general_solution}, we obtain \eqref{eq:deco-Helmholtz}. Using \eqref{eq:deco-T} and the harmonicity of $\Ti[\vec g]$ yields
  \begin{equation}
    \label{eq:div-v*}
    \begin{aligned}
      \div \vec v^*&=\Ti[\vec g]+\div \int_0^1 \frac{t|\vec x|^2}{2} \grad \Ti[\vec g](t\vec x) \, dt\\
      & =\Ti[\vec g]+\int_0^1 t\left(\vec x \cdot \grad \Ti[\vec g](t\vec x)+\dfrac{|\vec x|^2}{2} \div \grad \Ti[\vec g](t\vec x)\right)\, dt\\
      & =\Ti[\vec g]+\int_0^1 t\vec x \cdot \grad \Ti[\vec g](t\vec x)\, dt.
    \end{aligned}
  \end{equation}
  Note that $\div \vec v^*$ is harmonic if and only if the second term at the right-hand side of \eqref{eq:div-v*} is harmonic. Next, we will see that $\int_0^1 t\vec x \cdot \grad \Ti[\vec g](t\vec x)\, dt$ is harmonic. By the proof of \cite[Prop.\ 2.3]{DelPor2017}, $D(\nabla \Ti[\vec g](t\vec x) \vec x)=D \Ti[\vec g](t\vec x)$, which implies $\Delta (\nabla \Ti[\vec g](t\vec x) \vec x)=0$. As a consequence,
  \begin{equation}
    \int_0^1 t\nabla \Ti[\vec g](t\vec x)\vec x\, dt=-\int_0^1 t\vec x\cdot \grad \Ti[\vec x](t\vec x)\, dt-\U \Ti[\vec g](\vec x),
  \end{equation}
  is harmonic, which guarantees that its scalar part is also harmonic. The result now follows from \eqref{eq:div-v*}
\end{proof}

The similarity between the decomposition \eqref{eq:deco-Helmholtz} and the classical Helmholtz decomposition \eqref{eq:Helmholtz-R3} is evident when $\Omega$ is the entire three-dimensional space. One difference between these two solutions is that the vector Helmholtz potential $\vec v$ of \eqref{eq:Helmholtz-R3} is divergence-free, that is, $\div \vec v=0=T_{0,\R^3}[\vec g]$. In other words, the scalar part of the Teodorescu transform $T_{0,\R^3}$ defined in all $\R^3$ vanishes. On the other hand, $\div \vec v^*$ is harmonic in \eqref{eq:deco-Helmholtz}. In particular, the first term of $\vec v^*$ satisfies $\div L[\vec g]=T_{0,\Omega}[\vec g]$, so it is harmonic, and this property is enough to perform the monogenic completion process in the proof of Theorem \ref{th:div-curl}. The kernel of the scalar operator $\Ti$ will be studied in Proposition \ref{prop:kernel-T0}.  

\subsection{Right inverse of $\curl$\label{sec:right-inverse-curl}}

It has been proved \cite[Lem.\ 1]{Yoshida1990} that the $\curl$ operator has a compact inverse from $L_{\Sigma}^2(\Omega)$ to $L_{\Sigma, \curl}^2(\Omega)$. Here,
\begin{align}
  L_{\Sigma}^2(\Omega) & =\{\curl \vec u \colon \vec u\in W^{1,2}(\Omega), \div \vec u=0, \eta \times \vec u=0\}, \\
  L_{\Sigma, \curl}^2(\Omega) & = \{\vec u\in L_{\Sigma}^2(\Omega) \colon \curl \vec u\in L_{\Sigma}^2(\Omega)\}.
\end{align}
The proof is based on the following orthogonal decomposition of the space $L^2(\Omega)$:
\begin{equation}
  L^2(\Omega)=L_{\Sigma}^2(\Omega)\oplus \Ker \curl.
\end{equation}
Here, $\Ker \curl=\{\vec u\in L^2(\Omega) : \curl \vec u=0\}$. Let $\Sol^p(\Omega)$ be the subspace of $L^p$ consisting of all divergence-free functions (sometimes also called \emph {solenoidal vector fields}), and let $\Sol_0^p(\Omega)\subset \Sol^p(\Omega)$ be the subspace of solenoidal vector fields with vanishing normal trace. In notation,
\begin{align}
  \Sol^p(\Omega) & =\{ \vec u\in L^p(\Omega) \colon \div \vec u=0\}, \\
  \Sol_0^p(\Omega) & =\{ \vec u\in \Sol^p(\Omega) \colon \gamma_{\nrm}(\vec u)=\vec u|_{\partial\Omega}\cdot \eta=0\}.
\end{align}
In general, $L^2_{\Sigma}(\Omega)$ is a subset of $\Sol_0^2(\Omega)$. However, if $\Omega$ is simply connected, then the domain of definition of the compact inverse operator of $\curl$ studied in \cite{Yoshida1990} reduces to $L_{\Sigma}^2(\Omega)=\Sol_0^2(\Omega)$. We refer the reader to \cite[Th.\ 1]{Yoshida1990} for mode details on the self-adjointess of $\curl$ on $L_{\Sigma}^2(\Omega)$ and its spectral theory. It is important to mention that the $\curl$ operator is self-adjoint when it acts over vector fields with vanished tangential trace, its symmetry in this domain of definition is illustrated by the well-known Green's formula
\begin{equation}
  \int_{\Omega} (\curl \vec u\cdot \vec v-\vec u\cdot \curl \vec v) \, d\vec y=\int_{\partial\Omega} (\vec u|_{\partial\Omega}\times \vec v|_{\partial\Omega})\cdot \eta \, ds_{\vec y},
\end{equation}
where $\Omega$ has a sufficiently regular boundary $\partial\Omega$ and $\vec u, \vec v \in C^1(\overline{\Omega})$.
Additionally, the reader may check \cite{Hiptmair2012} and references therein for an analysis of self-adjoint $\curl$ operators.

It is worthwhile mentioning that an explicit expression for a right inverse of $\curl$ was reported as \cite[Th.\ 4.1]{DelPor2017}. The result was derived from the decomposition of \eqref{eq:general_solution} as the sum $\vec v_1+\vec v_2$, where the summands satisfy $\vec v_1=-\Tii[\vec g]$ and $\vec v_2=\Tiii[\vec g]-\U \Ti[\vec g]$ in $\Omega$. Moreover, the following identities are satisfied in $\Omega$:
\begin{equation}
  \begin{aligned}
    \div \vec v_1 & = g _0, \qquad & \div \vec v_2 = 0,\\
    \curl \vec v_1 & = 0,  \qquad & \curl \vec v_2 = \vec g.
  \end{aligned}
\end{equation}
Therefore, $-\Tii$ is a right inverse for the $\div$ operator, and 
\begin{equation}
  \label{eq:inverse-curl}
  R_{\Omega}[\vec g]=\Tiii[\vec g] - \U \Ti[\vec g]
\end{equation}
is a right inverse for the $\curl$ operator in the space of divergence-free functions. Moreover, $R_{\Omega}\colon \Sol^p(\Omega) \to \Sol^p(\Omega)$ is an invariant operator. 

\begin{proposition}
  \label{prop:boundedness}
  Let $\Omega$ be a star-shaped domain. The right inverse $R_{\Omega}$ for the curl operator \eqref{eq:inverse-curl} is bounded in $L^p(\Omega)$. Moreover, $R_{\Omega} : \Sol^p(\Omega)\to W^{p,\mbox{\scriptsize\rm div-curl}}(\Omega)$ is bounded.
\end{proposition}

\begin{proof}
  Recall that the Teodorescu operator $T_{\Omega}\colon L^p(\Omega)\to W^{1,p}(\Omega)$ is bounded \cite[Th.\ 8.4]{GuHaSpr2008}, and that $|\Sc a|\leq |a|$ and $|\Vec a|\leq |a|$ hold, for each $a\in \H$. As a consequence, 
  \begin{align}
    \|\Tiii[\vec g]\|_{L^p} & =\|\Vec T_{\Omega}[\vec g]\|_{L^p}\leq C_0 \|\vec g\|_{L^p}, \\
    \|\grad \Ti[\vec g]\|_{L^p} & \leq \|\Sc T_{\Omega}[\vec g]\|_{W^{1,p}}\leq C_1 \|\vec g\|_{L^p}. 
  \end{align}
  So, we only need to bound $\U \Ti[\vec g]$. Notice that
  \begin{align}\label{eq:bound-U}
     \nonumber|\U \Ti[\vec g](\vec x)|&\leq \int_0^1 |t\vec x| |\grad \Ti[\vec g](t\vec x)| \, dt \\
     \nonumber&\leq \dfrac{|\vec x|}{(q+1)^{1/q}} 
    \|\grad \Ti[\vec g]\|_{L^p} \\
    &\leq \dfrac{|\vec x|}{(q+1)^{1/q}} C_1 \|\vec g\|_{L^p}.
  \end{align}
  Integrating over $\Omega$, we readily obtain that $\|\U \Ti[\vec g]\|_{L^p}\leq C_{\Omega}\|\vec g\|_{L^p}$. As a consequence of \eqref{eq:inverse-curl}, it follows that $\|R_{\Omega}[\vec g]\|_{L^p} \leq C^*_{\Omega} \|\vec g\|_{L^p}$, as desired. Finally, the boundedness of $R _\Omega : \Sol^p(\Omega)\to W^{p,\mbox{\scriptsize\rm div-curl}}(\Omega)$ is a direct consequence of the fact that $\|R_{\Omega}[\vec g]\|_{W^{p,\mbox{\scriptsize div-curl}}}=\|R_{\Omega}[\vec g]\|_{L^p}+\|\vec g\|_{L^p}$.
\end{proof}

We can obtain now a right inverse operator for $\curl \curl$ by taking $g_0=0$ in \eqref{eq:deco-Helmholtz}. Indeed, let us define the operator $S_{\Omega}\colon \Sol^p(\Omega)\to W^{1,p}(\Omega)$ by
\begin{align}
  \label{eq:inverse-curlcurl}
  S_{\Omega}[\vec g]:=-L[\vec g]-\int_0^1 \frac{t|\vec x|^2}{2} \grad \Ti[\vec g](t\vec x) \, dt.
\end{align} 
As a consequence, given $\vec g\in \Sol^p(\Omega)$, there exists $S_{\Omega}[\vec g]\in W^{1,p}(\Omega)$ with the property that
\begin{equation}
  \begin{aligned}\label{eq:double-curl-inverse}
    \curl \curl \vec S_{\Omega}[\vec g] & =\vec g, \quad \text { in } \Omega, \\
    \Delta \div \vec S_{\Omega}[\vec g] & =0,\quad \text{ in }\Omega.
  \end{aligned}
\end{equation}
On the other hand, observe that our restriction to star-shaped domains in the construction of the operators $R_{\Omega}$ and $S_{\Omega}$ implies that the domains must be simply connected. 


\section{Results}\label{sec:BVP}
\subsection{Homogeneous Neumann condition\label{sec:Neumann BVP}}

In the present and the next subsection, we will consider respectively Neumann and Dirichlet BVP associated to the $\curl$ operator. In the present stage, we will require that the normal vector be defined almost everywhere at the boundary of $\Omega$. Obviously, this requirement is satisfied in the case when the domain is Lipschitz. For a fixed $\alpha > 0$ and $\vec x\in \partial\Omega$, the region of non-tangential approach with vertex at $\vec x$ is given by
\begin{align}
   \Gamma_{\alpha}(\vec x)=
  \left\{\vec y\in \Omega \colon |\vec x-\vec y|
  \leq (1+\alpha)\dist(\vec y,\partial\Omega)\right\}.
\end{align}
The function $\N_\alpha W\colon \partial\Omega \to [0,\infty]$ is the \textit{non-tangential maximal function} given by
\begin{equation}
  \label{eq:maximal-function}
  \N_{\alpha} w(\vec x)=\sup \left\{|w(\vec y)| :
	\vec x\in \Gamma_{\alpha}(\vec x)\right\}.
\end{equation}
When measuring the growth of $w$, the choice of $\alpha$ is largely irrelevant. 
In light of this remark, one simply writes $\N$ instead of $\N_\alpha$. Let $1<p<\infty$. The \textit{Hardy space} $H^p(\Omega)$ consists of all monogenic functions $w$ in $\Omega$ whose non-tangential maximal function
  $\N w$ belongs to $L^p(\partial\Omega,\R)$, that is,
\begin{align*}
   \|w\|_{H^p}:=\|\N w\|_{L^p(\partial\Omega)}<\infty.
\end{align*}

Following a compact-embedding argument, we can modify the operator $R_{\Omega}$ of \eqref{eq:inverse-curl} in order to obtain a compact operator. First, notice that the degree of freedom of the right inverse operator $R_{\Omega}$ is unique up to the sum of the gradient of a scalar function. That is, $R_{\Omega}[\vec g]+\nabla h$ is still a right inverse of $\curl$. However, we also want that the modified right inverse of $\curl$ continue leaving invariant the subspace $\Sol^p(\Omega)$ and have normal trace equal zero. To that end, let us define
\begin{align}
  \label{eq:modified-right-inverse}%
  R_{\Omega,\nrm}[\vec g]:=R_{\Omega}[\vec g]+\nabla h,
\end{align}
where $h$ is a scalar harmonic functions satisfying the Neumann problem
\begin{equation}
  \label{eq:Neumann-associated}
  \begin{aligned}
    \Delta h&=0, \quad \text{ in }\Omega,\\
    \nabla h|_{\partial\Omega} \cdot \eta&=-R_{\Omega}[\vec g]|_{\partial\Omega}\cdot \eta,\quad \text{ on }\partial\Omega.
  \end{aligned}
\end{equation}
Some works in the literature have studied this type of Neumann BVPs \cite {DahKen1987, FaJoLe1977, JeKe1981}. In the present report, we will employ the result in \cite{DahKen1987} for Lipschitz domains with connected boundary, which establishes that there exists a unique harmonic function $h$ in $\Omega$ which is unique up to constants, such that
\begin{align}
  \label{eq:bound-maximal-function}
  \|\N \nabla h\|_{L^p(\partial\Omega)}\leq C_{p,\Omega} \|R_{\Omega}[\vec g]|_{\partial\Omega}\cdot \eta\|_{L^p(\partial\Omega)}.
\end{align}
Here, $1<p<2+\epsilon$, and $\epsilon$ is the Lipschitz characteristic of the domain. 

As a consequence of these discussion, $R_{\Omega,\nrm} : \Sol^p(\Omega) \to \Sol_0^p(\Omega)$. Moreover, if $\Omega$ is a star-shaped domain with Lipschitz boundary, then $\vec w=R_{\Omega,\nrm}[\vec g]$ provides a unique weak solution to the first order system 
\begin{equation}
  \label{eq:Neumann-problem}%
  \begin{aligned}
    \curl \vec w&=\vec g, \quad \text{ in }\Omega,\\
    \vec w|_{\partial\Omega}\cdot \vec \eta&=0\quad \text{ on }\partial\Omega,
  \end{aligned}
\end{equation}
for all $\vec g\in \Sol^p(\Omega)$ with $1<p< 2+\epsilon$. In fact, a difference of two solutions of \eqref{eq:Neumann-problem} belongs to 
\begin{equation}
  \SI_{\nrm}(\Omega)=\{\vec u\in L^2(\Omega)\colon \div \vec u=0, \curl \vec u=0, \vec u|_{\partial\Omega}\cdot \eta=0\}, 
\end{equation}
which has finite dimension, and is trivial when $\Omega$ is simply connected \cite[Ch.\ 9, Cor.\ 2]{Dautray1985}. 


\begin{theorem}
  \label{th:right-inverse-compact}
  Let $\Omega$ be a star-shaped domain with Lipschitz boundary. The right inverse $R_{\Omega,\nrm}$ for the $\curl$ operator defined in \eqref{eq:modified-right-inverse}--\eqref{eq:Neumann-associated} is compact in $L^2(\Omega)$.
\end{theorem}

\begin{proof}
  We will check firstly that $R_{\Omega,\nrm}$ is bounded in $L^2(\Omega)$. By Proposition \ref{prop:boundedness}, it is enough to bound the last term $\nabla h$ in the expression \eqref{eq:modified-right-inverse}. Due to $\nabla h$ being monogenic and by \eqref{eq:bound-maximal-function}, it follows that $\nabla h\in H^2(\Omega)$. The equivalences provided by \cite[Thm.\ 4.1]{Mitrea1994} establish that $\nabla h=F_{\partial\Omega}[\nabla h]$. Using the continuity of the operator $F_{\partial\Omega}\colon L^p(\partial\Omega) \to L^p(\Omega)$, it follows that the next inequalities are satisfied:
  \begin{align}
    \|\nabla h\|_{L^2(\Omega)}
    &\leq \|F_{\partial\Omega}\| \|\nabla h\|_{L^2(\partial\Omega)}\leq \|F_{\partial\Omega}\| \|\N \nabla h\|_{L^2(\partial\Omega)}.
  \end{align}
  In this inequalities, the last one is a consequence of the definition of the non-tangential maximal function $\N$ in \eqref{eq:maximal-function}. On the other hand, the inequality \eqref{eq:bound-maximal-function} and the boundedness of the normal trace operator $\gamma\nrm \colon W^{2,\mbox{\scriptsize div}}(\Omega) \to H^{-1/2}(\partial\Omega)$ (see \cite[Ch.\ 9, Thm.\ 1]{Dautray1985}) guarantee that
  \begin{equation}
    \begin{aligned}	
      \|\nabla h\|_{L^2(\Omega)}&\leq \|F_{\partial\Omega}\| C_{\Omega} \|R_{\Omega}[\vec g]|_{\partial\Omega} \cdot \eta \|_{L^2(\partial\Omega)} \leq \|F_{\partial\Omega}\| C_{\Omega} \|R_{\Omega}[\vec g]|_{\partial\Omega}\cdot \eta\|_{H^{1/2}(\partial\Omega)}\\
      &\leq \|F_{\partial\Omega}\| C_{\Omega} \|\gamma\nrm\| \|R_{\Omega}[\vec g]\|_{W^{2,\mbox{\scriptsize div}}(\Omega)}.
    \end{aligned}
  \end{equation}
  Notice that the conclusion readily follows now from the compactness of the embedding of $W\nrm^{2,\div\text{-}\curl}(\Omega)$ into $L^2(\Omega)$ (see \cite[Th.\ 2.8]{ABDG1998} and \cite{Weber1980}).
\end{proof}

\begin{proposition}
  Let $\Omega$ be a star-shaped domain with Lipschitz boundary. The space $L^2(\Omega)$ allows the decomposition
  \begin{equation}
    L^2(\Omega)=\{\curl \vec u\colon \vec u\in \Sol_0^2(\Omega)\}\oplus \grad W_0^{1,2}(\Omega),
  \end{equation}
  under the scalar product $\langle \vec u,\vec v\rangle_{L^2}=\int_{\Omega}\vec u \cdot \vec v$.
\end{proposition}

\begin{proof}
  The proof follows from the decomposition $L^2(\Omega)=\Sol^2(\Omega)\oplus \grad W_0^{1,2}(\Omega)$ (see \cite{Temam1979}) as well as from the facts that $R_{\Omega,\nrm}\colon \Sol^2(\Omega)\to \Sol_0^2(\Omega)$ and $\vec g=\curl R_{\Omega,\nrm}[\vec g]$, for all $\vec g\in \Sol^2(\Omega)$.
\end{proof}

The \textit{single-layer potential} \cite{Mclean2000} is defined by 
\begin{align}\label{eq:single_layer_operator}
    M[\varphi](\vec x) &= \int_{\partial\Omega} 
    \frac{\varphi(\vec y )}{4\pi|\vec y -\vec x |} \,ds_{\vec y} ,\quad
    \vec x \in \R^3\setminus \partial\Omega.
\end{align}
 Meanwhile, the \textit{boundary single-layer operator} $\tr M$ is obtained by evaluating the integral in \eqref{eq:single_layer_operator} for $x\in\partial\Omega$. In such way, the single-layer potential $M$ is extended to all of $\R^3$. 

To further investigate the operators $R_{\Omega}$ and $R_{\Omega,\nrm}$, we will characterize the kernel of the component operator $\Ti$ involved in their constructions, as described by \eqref{eq:inverse-curl} and \eqref{eq:modified-right-inverse}, respectively. We will restrict the domain of $\Ti$ to the class of divergence-free functions $\Sol^p(\Omega)$. Under these circumstances, the following question emerged in \cite{DelPor2017} and was left as an open question in that report: under which conditions does the general solution \eqref{eq:general_solution} coincide with \eqref{eq:Helmholtz-R3}, which is the solution given by the classical Helmholtz Decomposition Theorem? An affirmative answer to that question is provided in the following result.

\begin{proposition}
  \label{prop:kernel-T0}%
  The kernel of the scalar integral operator $\Ti$ in $\Sol^2(\Omega)$ is the subspace $\Sol_0^2(\Omega)$.
\end{proposition}

\begin{proof}
  It was noted in \cite{DelPor2017} that the scalar component $T_{0,\Omega}$ can be written as $T_{0,\Omega}[\vec g]=M[\vec g|_{\partial\Omega} \cdot \eta]$, for all $\vec g\in \Sol^p(\Omega)$. By \cite[Th.\ 3.3]{Ver1984}, $\tr M \colon L^2(\partial\Omega) \to H^{1/2}(\partial\Omega)$ is invertible or, in the generalized sense, $\tr M : H^{-1/2}(\partial\Omega) \to H^{1/2}(\partial\Omega)$ (see \cite[Thm.\ 6.12]{Mclean2000}). Conclude that $\Ti[\vec g]=0$ if and only if $\vec g$ has zero normal trace. 
\end{proof}

\begin{remark}
  \label{rmk:reduction-inverse}%
  As a consequence of this proposition, if $\vec g\in \Sol_0^2(\Omega)$, then the general weak solution \eqref{eq:general_solution} reduces to 
  \begin{equation}
    \vec w=-\Tii[g_0]+\Tiii[\vec g]=\grad L[g_0]-\curl L[\vec g].
  \end{equation}
  Moreover, the right inverse of $\curl R_{\Omega}$ defined in \eqref{eq:inverse-curl} reduces to $R_{\Omega}=\Tiii$ (the Biot--Savart operator \eqref{eq:right-inverse-R3} over $\Omega$), and the modified right inverse $R_{\Omega,\nrm}$ reduces to $R_{\Omega,\nrm}=\Tiii+\nabla h$, with $h$ the solution of the Neumann problem \eqref{eq:Neumann-associated}.
\end{remark}

\subsection{Homogeneous Dirichlet condition\label{sec:Dirichlet BVP}}

In the present stage of our work, we are interested in the analysis of the BVP with homogeneous Dirichlet condition
\begin{equation}
  \label{eq:Dirichlet-curl}
  \begin{aligned}
    \curl \vec w=\vec g, \quad \text{ in } \Omega,\\
    \vec w|_{\partial\Omega}=0, \text{ on } \partial\Omega.
  \end{aligned}
\end{equation}
By the well-known Helmholtz decomposition for vector fields in simply connected domains, it follows that
\begin{equation}
  L^2(\Omega)=\{\curl \vec w \colon \vec w\in W_0^{1,2}(\Omega)\}\oplus \{\nabla h\colon h\in h\in W^{1,2}(\Omega)\}.
\end{equation} 
It was noted in \cite{Griesinger1990} that \eqref{eq:Dirichlet-curl} has a unique solution when $\vec g\in\{\nabla h\colon h\in h\in W^{1,2}(\Omega)\}$. It is important to point out that the authors of \cite[Cor.\ 8']{Brezis2007} proved a result on the existence of a right inverse of $\curl$. They also established some estimates using results on differential forms with coefficients in Sobolev spaces.
 
We claim that $\vec g\in \Sol^p(\Omega)$ in \eqref{eq:Dirichlet-curl} will necessarily have vanishing normal trace. Indeed, notice that
\begin{equation}
  0=\Div(\vec w|_{\partial\Omega}\times \eta)=\curl \vec w|_{\partial\Omega}\cdot \eta=\vec g|_{\partial\Omega}\cdot \eta,
\end{equation}
where $\Div$ is the surface divergence. Moreover,

\begin{proposition}
  \label{prop:vanishing-trace-normal}
  If $\vec g\in \Sol^2(\Omega)$ and $\vec w\in W^{2,\mbox{\scriptsize div}}(\Omega)$ be a solution of the system \eqref{eq:Dirichlet-curl}, then $\vec g\in \Sol_0^2(\Omega).$
\end{proposition}

\begin{proof}
  Let $\vec w\in W^{2,\mbox{\scriptsize div}}(\Omega)$ is a solution of $\curl \vec w=\vec g$ with $\vec w|_{\partial\Omega}=0$. Friedrichs' inequalities imply that $\vec w\in W^{1,2}(\Omega)$ (see \cite[Ch.\ 9, Cor.\ 1]{Dautray1985}). The Borel--Pompeiu formula \cite{GuHaSpr2008} and the decomposition \eqref{eq:decomposition_T} yield that 
  \begin{equation}
    \vec w=T_{\Omega}[D\vec w]=T_{\Omega}[-\div \vec w+\vec g]=\Ti[\vec g]-\Tii[\div \vec w]+\Tiii[\vec g].
  \end{equation} 
  As a consequence, the scalar part on the right-hand side of this identity vanishes. Proposition \ref{prop:kernel-T0} shows now that $\Ti[\vec g]=0$ if and only if $\vec g\in \Sol_0^2(\Omega)$, as desired.
\end{proof}

Using some properties of the Teodorescu transform $T_{\Omega}$, it is possible to prove that this class of solutions vanishes not only at the boundary but also in the entire \emph {exterior of the domain} which is the set $\Omega^{-}=\R^3\setminus \overline{\Omega}$. In other words,  we have the following result.

\begin{proposition}
  Let $\Omega$ be a star-shaped domain. Let $\vec g\in \Sol^2(\Omega)$. Then the solutions of \eqref{eq:Dirichlet-curl} belonging to $W^{2,\mbox{\scriptsize div}}(\Omega)$ vanishes in the exterior domain $\Omega^{-}$.
\end{proposition}

\begin{proof}
  Propositions \ref{prop:kernel-T0} and \ref{prop:vanishing-trace-normal} yield $\Ti[\vec g]=0$. Using then the Borel--Pompeiu formula, we obtain that $\vec w=T_{\Omega}[D\vec w]$ and $\vec w\in W\nrm^{2,\div\text{-}\curl}(\Omega)$. Use now the fact that $DT_{\Omega}=T_{\Omega}D$ holds in $W_0^{1,2}(\Omega)$ and that the Teodorescu transform $T_{\Omega}$ is monogenic in $\Omega^{-}$ (see \cite[Prop.\ 8.1]{GuHaSpr2008}) to establish that $\vec w=T_{\Omega}[D\vec w]=DT_{\Omega}[\vec w]=0$ in $\Omega^-$.
\end{proof}

The novelty of this right inverse operator $R_{\Omega}$ is precisely the term that involves the radial operator $\U$ acting on $\Ti$. However, Propositions \ref{prop:vanishing-trace-normal} and \ref{prop:kernel-T0} show that $\U\Ti[\vec g]$ vanishes. 

On the other hand, if we modify the operator $R_{\Omega}$ to find a solution of \eqref{eq:Dirichlet-curl} as we did in Section \ref{sec:Neumann BVP}, the modified operator will be similar to that constructed in \cite[Cor.\ 2.3]{Borchers1990}. Using that $T_{\Omega}$ is monogenic in $\Omega^{-}$, that $\Ti[\vec g]=0$ in $\Omega$ and the maximum Principle, we obtain that $\Ti[\vec g]\equiv 0$ in all $\R^3$. So $\curl \Tiii[\vec g]=0$ in $\Omega^-$. In the following, we let $p^*\in W^{2,p}_{\text{loc}}(\Omega^-)$ be the \emph {antigradient} of $\Tiii[\vec g]$, that is, let $\Tiii[\vec g]=\nabla p^*$ in $\Omega^{-}$. Let us define
\begin{align}
  \label{eq:Dirichlet-inverse}
  R_{\Omega,0}[\vec g]:=\Tiii[\vec g]-\nabla p, 
\end{align}
where $p$ is the biharmonic function satisfying the Dirichlet boundary value problem (see \cite{Simader1972})
\begin{equation}
  \label{eq:bvp-biharmonic}
  \begin{aligned}
    \Delta^2 p&=0,\quad \text{ in }\Omega,\\
    \nabla p|_{\partial\Omega}&=\nabla p^*|_{\partial\Omega},\quad \text{ on }\partial\Omega.
  \end{aligned}
\end{equation}
Moreover, $p\in W^{2,p}(\Omega)$.

\begin{proposition}
  Let $\Omega$ be a $C^2$ bounded simply connected domain with connected boundary. Then the operator $R_{\Omega,0}\colon \Sol_0^p(\Omega)\to W_0^{1,2}(\Omega)$ defined in \eqref{eq:Dirichlet-inverse} and \eqref{eq:bvp-biharmonic} is a right inverse of $\curl$ with vanishing Dirichlet condition. In other words, $\vec w=R_{\Omega,0}[\vec g]$ provides a weak solution of \eqref{eq:Dirichlet-curl}, with $\Delta \div \vec w=0$.\qed
\end{proposition}

\section{Beltrami fields\label{sec:Beltrami-fields}}
This section is devoted to the construction of Beltrami fields through an uniformly convergent Neumann series in terms of the inverse curl operator $R_{\Omega}$ analyzed in this work. It is worth mentioning that quaternionic analysis techniques have been used previously for the generation of Beltrami fields, a recent example is \cite{kravchenko2021complete}. 

Recall that a \textit{Beltrami field} in $\Omega$ is a vector field $\vec w\colon \Omega \to \R^3$ satisfying the equation
\begin{align}
  \label{eq:Beltrami-field}
  \curl \vec w=\alpha_0 \vec w, \quad \text{ in }\Omega,
\end{align}
where the potential $\alpha_0$ is a real-valued function (see \cite{Chandrasekhar1957,Jette1970}). Observe that if $\alpha _0 = c _0 \in \R$ is a constant, then $\div \vec w = 0$. On the other hand, if $\alpha_0$ is a differentiable function, then the compatibility conditions transform into
\begin{align}
   \div(\alpha_0\vec w)=\nabla \alpha_0\cdot \vec w+\alpha_0 \div \vec w=0.
\end{align}

Let $\alpha_0\in L^{\infty}(\Omega)$. By Proposition \ref{prop:boundedness}, the operator $R_{\Omega}$ is bounded in $L^p(\Omega)$. Let $\|R_{\Omega}\|$ its norm operator from $L^p(\Omega)$ to itself. If $\|R_{\Omega}\alpha_0 I\|<1$ or $\|\alpha_0\|_{L^{\infty}}< 1/\|R_{\Omega}\|$, then $I-R_{\Omega}\alpha_0 I$ has a bounded inverse. Moreover, the Neumann series of the operator $(I-R_{\Omega}\alpha_0 I)^{-1}$ (see \cite[Th.\ 1.3]{Kub2012} and  \cite[Ex.\ 4.15]{GuHaSpr2008}) is given by the following uniformly convergent series:
\begin{align}\label{eq:Neumann-series}
  (I-R_{\Omega}\alpha_0 I)^{-1}=\sum_{k=0}^{\infty} (R_{\Omega}\alpha_0 I)^k,
\end{align}
where $R_{\Omega}\colon \Sol^p(\Omega)\to \Sol^p(\Omega)$ is a right inverse operator of $\curl$ previously defined in \eqref{eq:inverse-curl}.
Let us denote the class of irrotational vector fields as $\Irr^p(\Omega)=\{\vec u\in L^p(\Omega) : \curl\vec u=0\}.$
The next result shows a way to generate Beltrami fields with $\alpha_0\in \R$ constant, using Neumann expansion series.

\begin{theorem}
  \label{th:construction-Beltrami-constant}
  Let $\Omega$ be a star-shaped domain. Let $1<p<\infty$, $\vec g\in \Sol^p(\Omega)\cap \Irr^p(\Omega)$ and $\alpha=\alpha_0\in \R$. If $\|R_{\Omega}\alpha_0 I\|<1$ or $|\alpha_0|< 1/\|R_{\Omega}\|$, then
  \begin{align}\label{eq:Neumann-serie-solution}
    \vec w=\sum_{k=0}^{\infty} (R_{\Omega}\alpha_0 I)^k[\vec g]=\vec g+R_{\Omega}[\alpha_0\vec g]+R_{\Omega}[\alpha_0R_{\Omega}[\alpha_0\vec g]]+\ldots
  \end{align} 
  is a Beltrami field in $\Omega$.
\end{theorem}

\begin{proof}
  Beforehand, notice that the iterated application of the operator \\$R_{\Omega}\colon \Sol^p(\Omega)\to \Sol^p(\Omega)$ is feasible in view that $\div(\alpha_0\vec g)=\alpha_0\div \vec g=0$ and $\div(\alpha_0 R_{\Omega}[\cdot])=\alpha_0 \div R_{\Omega}[\cdot]=0$. Applying the curl operator to both ends of the Neumann series \eqref{eq:Neumann-serie-solution}, we obtain that
  \begin{equation}
    \begin{aligned}
      \curl \vec w&=\curl \vec g+\curl R_{\Omega}[\alpha_0\vec g]+ \curl R_{\Omega}[\alpha_0R_{\Omega}[\alpha_0\vec g]]+\ldots\\
      &=\alpha_0 \vec g+\alpha_0R_{\Omega}[\alpha_0\vec g]+ \alpha_0R_{\Omega}[\alpha_0R_{\Omega}[\alpha_0\vec g]]+\ldots\\
      &=\alpha_0 \vec w.
    \end{aligned}
  \end{equation}
  It follows that $\curl \vec w=\alpha_0\vec w$ in $\Omega$ and $\div \vec w=\div \vec g=0$, as we desired.
\end{proof}


\begin{example}\label{ex:constant}
  Let $\Omega$ be the unit ball in $\R^3$, and let $\alpha_0=c_0 \in \R$ and $\vec g=\vec c\in \R^3$ satisfy the hypotheses of Theorem~\ref{th:construction-Beltrami-constant}. The explicit formulas for the Teodorescu transform obtained in \cite[App.\ A]{GuSpr1997} read as follows:
  \begin{equation}
    T_{\Omega}[\vec c](\vec x)=T_{\Omega}[1]\vec c=(1/3)(\vec x\cdot \vec c-\vec x\times \vec c),
  \end{equation}
  Using these expressions, it is easy to check that  
  \begin{align}
    \nonumber R_{\Omega}[\vec c] &=\Tiii[\vec c]-\U \Ti[\vec c] =-\dfrac{1}{3} \vec x\times \vec c-\Vec \left(\int_0^1 t\vec x \dfrac{1}{3} \nabla_{\vec x}(\vec x\cdot \vec c)\, dt\right) \\
    &= -\dfrac{1}{3} \vec x\times \vec c- \dfrac{1}{6} \vec x\times \vec c=-\frac{1}{2} \vec x\times \vec c.
  \end{align}
  As a consequence, the modified right inverse of $\curl$ has the same expression. That is, $R_{\Omega,\eta}[\vec c]=R_{\Omega}[\vec c]=-(1/2)\vec x\times \vec c$, due to $\eta(\vec x)=\vec x$. Observe that we have constructed implicitly a general solution $\vec v=-\frac{1}{2} \vec x\times \vec c+\nabla h$ for the div-curl system $\div \vec v=0$, $\curl \vec v=\vec c$. Here, $h$ is an arbitrary harmonic function (we refer to \cite[Ex.\ 4.5]{DelPor2017} for another example with non-constant $\curl$). To compute $T_{\Omega}[\vec x\times \vec c]$ in this iterative process, observe that
  \begin{equation}
    T_{\Omega}[\vec x\times \vec c]=T_{\Omega}[\vec x]\vec c+T_{\Omega}[\vec c\cdot \vec x] =\dfrac{1}{2}(|\vec x|^2-1)\vec c+\Tii[\vec x\cdot \vec c],
  \end{equation}
  is purely vectorial. According to Proposition \ref{prop:kernel-T0}, $\Ti[\vec x\times \vec c]=0$. Using the Borel--Pompeiu formula, it is possible to check that $\Tii[\vec x\cdot \vec c]=(1/2)(c_1(x_1^2-1),c_2(x_2^2-1),c_3(x_3^2-1))$. Thus,
  \begin{equation}
    \begin{aligned}
      R_{\Omega}[R_{\Omega}[\vec c]]&=-\dfrac{1}{2}R_{\Omega}[\vec x\times \vec c]=\Tiii[\vec x\times \vec c]\\
      &=-\dfrac{1}{4} (|\vec x|^2-1)\vec c-\dfrac{1}{4}(c_1(x_1^2-1),c_2(x_2^2-1),c_3(x_3^2-1)).
    \end{aligned}
  \end{equation}
  Finally, the first few terms of the Neumann series of the Beltrami field $\vec w$ constructed in this example are given by
  \begin{equation}
    \vec w(\vec x)=\vec c-\dfrac{c_0}{2}\vec x\times\vec c-\dfrac{c_0^2}{4}\left((|\vec x|^2-1)\vec c+(c_1(x_1^2-1),c_2(x_2^2-1),c_3(x_3^2-1))\right)+\cdots 
  \end{equation}
\end{example} 

We will describe next a method to generate Beltrami fields subject to a Neumann condition. To that end, we will use the modified right inverse of $\curl$ $R_{\Omega,\nrm}\colon \Sol^p(\Omega)\to \Sol_0^p(\Omega)$ defined in \eqref{eq:modified-right-inverse}. This operator is not only bounded operator in $L^p(\Omega)$, but it is also compact (see Theorem \ref{th:right-inverse-compact}).

\begin{proposition}
  \label{prop:BVP-Beltrami}
  Let $\Omega$ be a star-shaped domain with Lipschitz boundary. Let $\vec g\in \Sol^2(\Omega)\cap \Irr^2(\Omega)$. Suppose that $\|R_{\Omega,\nrm}\alpha_0 I\|<1$ or $|\alpha_0|< 1/\|R_{\Omega,\nrm}\|$. Then
  \begin{align}
    \label{eq:Neumann-serie-solution2}
    \vec w=\sum_{k=0}^{\infty} (R_{\Omega,\nrm}\alpha_0 I)^k[\vec g]=\vec g+R_{\Omega,\nrm}[\alpha_0\vec g]+R_{\Omega,\nrm}[\alpha_0R_{\Omega,\nrm}[\alpha_0\vec g]]+\ldots
  \end{align} 
  is a Beltrami field in $\Omega$ satisfying the Neumann boundary condition $\vec w|_{\partial\Omega}\cdot \eta=a_0$ if and only if $\vec g=\nabla h$, where $h$ a solution of the Neumann BVP
  \begin{equation}
    \begin{aligned}
      \Delta h&=0,\quad \text{ in }\Omega,\\
      \nabla h|_{\partial\Omega}\cdot \eta&=a_0,\quad \text{ on }\partial\Omega.
    \end{aligned}
  \end{equation}
\end{proposition}
Observe that this construction of Beltrami fields relies on the condition $\|R_{\Omega}\alpha_0 I\|<1$ (or $|\alpha_0|< 1/\|R_{\Omega}\|$). The authors of this manuscript are aware that a sharper bound in terms of the operator norm of $R_{\Omega}$ is needed. In the following, we will give a bound for $|\alpha_0|$ which depends on $\Omega$ and the operator norm of the Teodorescu operator $T_{\Omega}$. Indeed, it is easy to compute an upper bound for the $\U\circ \Ti$ operator. By \eqref{eq:bound-U}, we readily obtain
\begin{equation}
|U_{\Omega}[\Ti[\vec g]](\vec x)|\leq \dfrac {\text{diam }(\Omega)}{(q+1)^{1/q}}\|\grad \Ti[\vec g]\|_{L^p}, \quad \forall \vec x\in \Omega.
\end{equation}
Therefore, integrating over $\Omega$ and using the boundedness of the Teodorescu transform from $L^p(\Omega)$ to $W^{1,p}(\Omega)$, we have 
\[
   \|U_{\Omega}[\Ti[\vec g]]\|_{L^p} \leq \dfrac{1}{(q+1)^{1/q}}\text{Vol }(\Omega) \, \text{diam }(\Omega) \|T_{\Omega}\|_{L^p\to W^{1,p}}\, \|g\|_{L^p},  \quad \forall \vec g\in L^p(\Omega),
\]
where $\|T_{\Omega}\|_{L^p\to W^{1,p}}$ is the operator norm from $L^p(\Omega)$ to $W^{1,p}(\Omega)$.
Consequently, $\dfrac{\|T_{\Omega}\|_{L^p\to W^{1,p}}}{(q+1)^{1/q}}\text{Vol }(\Omega) \, \text{diam }(\Omega)$ is an upper bound for $\|\U\circ \Ti\|$ from $L^p(\Omega)$ to itself. Due to $R_{\Omega}[\vec g]=\Tiii[\vec g]-\U[\Ti[\vec g]]$, then 
\begin{equation}
  \|R_{\Omega}\|\leq 2\max \left\{ \|T_{\Omega}\|_{L^p}, \dfrac{\|T_{\Omega}\|_{L^p\to W^{1,p}}}{(q+1)^{1/q}}\text{Vol }(\Omega) \, \text{diam }(\Omega)\right\}.
\end{equation}
Taking 
\begin{equation}
|\alpha_0|< 2\min\left\{ \dfrac{1}{\|T_{\Omega}\|_{L^p}}, \dfrac{(q+1)^{1/q}}{\|T_{\Omega}\|_{L^p\to W^{1,p}}\,\text{Vol }(\Omega) \, \text{diam }(\Omega)} \right\}, 
\end{equation}
we ensure that the norm of the composition of operators $R_{\Omega}\alpha_0 I$ is strictly less than one, as required by the Neumann series \eqref{eq:Neumann-series}. 

It is easy to compute an upper bound for the norm of the $\U^*$ operator is easy to compute. By \eqref{eq:bound-U}, we readily obtain
\begin{equation}
|U_{\Omega}[\vec g](\vec x)|\leq \dfrac {\text{diam }(\Omega)}{(q+1)^{1/q}}\|\vec g\|_{L^p}, \quad \forall x\in \Omega.
\end{equation}
Therefore, integrating over $\Omega$:
\[
   \|U_{\Omega}[\vec g]\|_{L^p} \leq \dfrac{1}{(q+1)^{1/q}}\text{Vol }(\Omega) \, \text{diam }(\Omega) \|g\|_{L^p},  \quad \forall \vec g\in L^p.
\]
Consequently, taking $|\alpha_0|\leq \dfrac{(q+1)^{1/q}}{\text{Vol }(\Omega) \, \text{diam }(\Omega)}$, we ensure that the norm of the composition of operators $U_{\Omega}\alpha_0 I$ is strictly less than one, as required by the Neuman series. 

\section{Vekua-type problems and its applications to the inhomogeneous Maxwell equations}\label{sec:Maxwell}

\subsection{The operator $D-\vec \alpha$}\label{sec:The operator D+alpha}\label{sec:operator D-vec-alpha}

The purpose of this section is to analyze the system 
\begin{align}
  \label{eq:D+alpha}
  (D-\vec \alpha)\vec w=g, 
\end{align} 
where $g\in L^p(\Omega)$ and $D$ is the Moisil--Teodorescu operator defined in \eqref{eq:Moisil-Teodorescu}. The identity $D\vec w=-\div \vec w+\curl \vec w$ guarantees that \eqref{eq:D+alpha} is equivalent to the following div-curl system:
\begin{equation}
  \label{eq:equiv1-D+alpha}
  \begin{aligned}
    \div \vec w -\vec \alpha \cdot \vec w & =-g_0,\\
	  \curl \vec w-\vec \alpha\times\vec w & =\vec g.
	\end{aligned}
\end{equation}
Taking the divergence in the second equation and using that $\curl \vec \alpha=0$, we obtain
\begin{equation}
  \div (\vec g)=\div(\vec w\times \vec \alpha)=\vec \alpha \cdot \curl \vec w=\vec \alpha \cdot (\vec g-\vec w\times \vec \alpha)=\vec \alpha \cdot \vec g.
\end{equation}


Let $\Omega$ be star-shaped and take $a=(a_1,a_2,a_3)\in \Omega$. Define the \textit{antigradient} operator $\mathcal{A}$ by
\begin{align}\label{eq:antigradient}
  \A[\vec u](\vec x) &= 
    \int_{a_1}^{x_1}{u_1(t,a_2,a_3) \,dt} +
    \int_{a_2}^{x_2}{u_2(x_1,t,a_3) \, dt}+ \int_{a_3}^{x_3}{u_3(x_1,x_2,t) \,dt},
\end{align}
where $\vec u$ is any vector field in the class $\Irr ^p(\Omega)$. Since $|\A[\vec u](\vec x)|\leq \max_{i} |x_i-a_i| \max_{i} \|u_i\|_{L^p(\Omega)}$ and $\nabla \A[\vec u]=\vec u$, we obtain that $\A : \Irr^p(\Omega)\to W^{1,p}(\Omega)$, for all $1\leq p\leq \infty$.

\begin{remark}
  \label{rmk:antigradient}
  If $\Omega$ is a star-shaped domain and $\vec \alpha\in \Irr^p(\Omega)$, then it is always possible to construct a positive scalar function $\varphi\in W^{1,p}(\Omega)$ such that $\vec \alpha=\nabla \varphi/\varphi$, namely, 
  \begin{equation}\label{eq:varphi-construction}
    \log \varphi=\A[\vec \alpha].
  \end{equation} 
\end{remark}

Another important feature about this class of irrotational vector fields was illustrated by the example in \cite{Sprossig1993}. In that work, the authors considered
\begin{equation}
  \begin{aligned}
    \varphi(\vec x) & = \text{exp}\left(\frac{1}{\vec x-\vec x^i}\right), \\
    D-\nabla \varphi/\varphi & =D+\left(\frac{\vec x-\vec x^i}{|\vec x-\vec x^i|^3}\right),
  \end{aligned}
\end{equation}
where $\vec x^{(i)}\in \R^3\setminus \overline{\Omega}$ is dense in a closed surface outside $\Omega$. Since the system $\left\{\frac{\vec x-\vec x^i}{|\vec x-\vec x^i|^3}\right\}_{i}$ is complete in $L^2(\Omega)\cap \text{Ker }D$ (see \cite[Th.\ 10.4]{GuHaSpr2008}), it follows that this class of irrotational vector fields is quite large.

On the other hand, the analysis of the operator $D-\vec \alpha$ was trivialized in \cite{Sprossig1993} assuming that $\vec \alpha$ has the form $\nabla \varphi/\varphi$. Indeed, in that case, the factorization $D-\vec \alpha=\varphi D \varphi^{-1}$ holds. This factorization allowed to find a straightforward right inverse of $\varphi D\varphi^{-1}$ in terms of the classical Teodorescu transform $T_{\Omega}$ \cite{Sprossig1993, Sprossig2001}: if $\varphi^{-1}w\in L^p(\Omega)$, then
\begin{align}\label{eq:varphi-Teodorescu}
  T_{\Omega,\varphi}[w](\vec x):=-\int_{\Omega}  \varphi(\vec x)E(\vec y-\vec x) \varphi^{-1}(\vec y) w(\vec y) \, d\vec y=\varphi(\vec x)T_{\Omega}\left[\frac{w}{\varphi^{-1}}\right](\vec x).
\end{align}
Therefore, an immediate consequence is that (see \cite[Lemma\ 1]{Sprossig1993})
\begin{align}\label{eq:right-inverse}
   \left(D-\frac{\grad \varphi}{\varphi}\right)T_{\Omega,\varphi}[w](\vec x)=\left\{
   \begin{array}{ccc}
      w(\vec x), &\text{  }\vec x\in \Omega,\\
      0, &\text{  } \vec x\in \R^3\setminus\overline{\Omega}.
   \end{array}
\right. 
\end{align}
We will see how the factorization $\varphi D\varphi^{-1}$ allows us to give an explicit solution to the system \eqref{eq:D+alpha} in terms of our solution to the div-curl system \eqref{eq:general_solution}. 
\begin{theorem}
  \label{th:solution-D+alpha}
  Let $\Omega$ be a star-shaped domain, and let $\vec \alpha\in L^{\infty}(\Omega)$ be such that $\curl \vec \alpha=0$. Suppose that $g\in L^p(\Omega)$ satisfies the compatibility condition $\div \vec g=\vec \alpha\cdot \vec g$. Then a weak solution of \eqref{eq:D+alpha} is given by
  \begin{align}
    \vec w=\varphi \left(-\Tii\left[\frac{g_0}{\varphi}\right]  +  R_{\Omega}\left[\frac{\vec g}{\varphi}\right] + \nabla h\right),
  \end{align}
  where $\varphi$ is constructed as in \eqref{eq:varphi-construction}, $R_{\Omega}$ is the right inverse of the $\curl$ operator defined in \eqref{eq:inverse-curl} and $h$ is an arbitrary harmonic function. 
\end{theorem}

\subsection{The operator $D+M^{\vec \alpha}$ \label{sec:operator-D+M}}
Let $M^{\alpha}$ be the right-hand side multiplication operator by the function $\alpha$, usually we will employ bounded functions. 
To start with, notice that we can readily observe that we can develop a solution method of the equation 
\begin{align}\label{eq:D+M}
  (D+M^{\vec \alpha})\vec w=g, \quad \text{ with }  g\in L^p(\Omega),
\end{align}
for the class of $p$-integrable irrotational vector fields $\vec \alpha$. That is, $\vec \alpha\in \Irr^p(\Omega)$. This method hinges on the combination of the theory developed here to solve the div-curl system in star-shaped domain in $\R^3$ and some variational methods employed in the theory of elliptic partial differential equations. Unlike the operator $D+\vec \alpha$ for which there exists a complete functional theory if $\vec \alpha=\nabla \varphi/\varphi$ (generalized Teodorescu and Cauchy operators, Borel--Pompeiu formula, Plemelj--Shokotski, etc.), there is no integral operator theory for the operators $D+\alpha_0$ and $D+M^{\vec \alpha}$, when $\alpha_0$ and $\vec \alpha$ are non-constant scalar and vector functions respectively.
 
Notice that the system \eqref{eq:D+M} is equivalent to the following type div-curl system:
\begin{equation}
  \label{eq:equivalent-D+M}
  \left\{ \begin{aligned}
    \div \vec w+\vec \alpha \cdot \vec w & =-g_0, \\
	  \curl \vec w+\vec w\times\vec \alpha & =\vec g.
	\end{aligned} \right.
\end{equation}
Comparing the systems \eqref{eq:equivalent-D+M} and \eqref{eq:equiv1-D+alpha} we can observe that the second equations in both systems corresponding to the vector part is equal. Therefore, under the hypothesis $\curl \vec \alpha=0$, we obtain the same \textit{compatibility condition} as that obtained for $D-\vec \alpha$ in Section \ref{sec:operator D-vec-alpha}, namely, $\div \vec g=\vec \alpha\cdot \vec g$.

\begin{theorem}
  \label{th:solution-D+M}
  Let $\Omega$ be a star-shaped domain, and let $\vec \alpha\in L^{\infty}(\Omega)$ be such that $\curl \vec \alpha=0$. Suppose that $g\in L^2(\Omega)$ satisfies the compatibility condition $\div \vec g=\vec \alpha\cdot \vec g$. Then a weak solution of \eqref{eq:D+M} is given by
  \begin{align}
    \vec w=\varphi R_{\Omega}\left[\frac{\vec g}{\varphi}
		\right]-\varphi \nabla w_0+\vec u,
  \end{align}
  where $\varphi$ is constructed as in \eqref{eq:varphi-construction}, $w_0$ is a solution of the conductivity equation $\div(\varphi^2 \nabla w_0)=\varphi g_0+2\nabla\varphi \cdot \vec w^*$, $\vec w^*=\varphi R_{\Omega}[\vec g/\varphi]$, and $\vec u$ is an arbitrary vector field belonging to the kernel of $D+M^{\vec \alpha}$. 
\end{theorem}

\begin{proof}
  Using Remark \ref{rmk:antigradient}, there exists a scalar function $\varphi\in W^{1,\infty}(\Omega)$ such that $\vec \alpha=\nabla \varphi/\varphi$. This implies that the equivalent system \eqref{eq:equivalent-D+M} can be expressed alternatively as
  \begin{equation}
    \label{eq:system-2}
    \begin{aligned}
      \frac{1}{\varphi} \div (\varphi \vec w) & =-g_0, \\
      \varphi \curl \left(\frac{\vec w}{\varphi}\right) & =\vec g.	
    \end{aligned}
  \end{equation}
  Notice that the right inverse of $\curl$ $R_{\Omega}\colon \Sol^p(\Omega)\to \Sol^p(\Omega)$ in \eqref{eq:inverse-curl} allows us to obtain a solution of the second equation of \eqref{eq:system-2}, though some adjustments are required to satisfy also the first. To that end, let
  \begin{equation}
    \vec w^* =\varphi R_{\Omega}\left[\frac{\vec g}{\varphi}\right].
  \end{equation}
  The application of the operator $R_{\Omega}$ is well-defined since $\div (\vec g/\varphi)=\varphi^{-1}(\div \vec g-\vec \alpha \cdot \vec g)=0$ holds by the compatibility condition. As a consequence, $\div(\vec w^*/\varphi)=0$ and $\curl(\vec w^*/\varphi)=\vec g/\varphi$. Let us define
  \begin{equation}
    \varphi \vec w:=\varphi\vec w^*-\varphi^2 \nabla w_0,
  \end{equation}
  where $w_0$ is a solution of the elliptic conductivity equation 
  \begin{align}\label{eq:conductivity}
    \div(\varphi^2 \nabla w_0)=\varphi g_0+2\nabla\varphi \cdot \vec w^*, \quad \text{ in }\Omega.
  \end{align}
  The existence of a solution of \eqref{eq:conductivity} is well-known, and it is based on the use of variational methods (see \cite[Theorem\ 4.1]{Isakov1998} and \cite[Theorem\ 10]{Mikhailov1978}). More precisely, we need to minimize the following functional $\epsilon$ in $W^{1,2}(\Omega)$:
  \begin{equation}
    \epsilon[u_0]=\int_{\Omega} \varphi^2 \nabla u_0\cdot \nabla u_0 + 2(\varphi g_0+2\nabla\varphi \cdot w^*) u_0, \quad \forall u_0\in W^{1,2}(\Omega).
  \end{equation}
  Moreover, the uniqueness of the minimum is guaranteed under some boundary Dirichlet condition. Without loss of generality, let us suppose that $w_0|_{\partial\Omega}=0$. We only need to verify that $\varphi g_0+2\nabla\varphi \cdot w^*\in L^2(\Omega)$, but this follows from the fact that $\varphi\in W^{1,\infty}(\Omega)$, $g=g_0+\vec g\in L^2(\Omega)$ and $R_{\Omega}[\vec g/\varphi]\in L^2(\Omega)$. We conclude that $\vec w$ satisfies \eqref{eq:D+M}. 
\end{proof}
\subsection{Time-indedendent Maxwell system in inhomogeneous media}
Finally, let us consider the \textit{Maxwell system in inhomogeneous media}, where the permittivity and permeability are bounded scalar functions in $\Omega$, $\epsilon=\epsilon(\vec x)$, $\mu=\mu(\vec x)\in L^{\infty}(\Omega)$. More precisely, assume that the following hold:
\begin{equation}
  \label{eq:Maxwell-1}
  \begin{aligned}
    \curl \vec{H}&=\epsilon\partial_t \vec{E}+\vec{j},\quad & \div(\mu\vec{H})=0,\\ 
    \curl\vec{E}&=-\mu\partial_t \vec{H},\quad &\div (\epsilon\vec{E})=\rho. 
  \end{aligned}
\end{equation}
Here, the charge and the current densities are related by the identity $\rho=\frac{1}{i\omega}\operatorname{div}\vec{j}$. Following \cite{Krav2002} and \cite[Ch.\ 4]{Krav2003}, if the electric and magnetic fields are time-independent, then we can rewrite \eqref{eq:Maxwell-1} in terms of the $D+M^{\vec \alpha}$ operator as
\begin{equation}
  \label{eq:Maxwell-2}
  \begin{aligned}
    (D+M^{\vec \epsilon}) \vec{\mathcal{E}} & =-\frac{\rho}{\sqrt{\epsilon}},\\
    (D+M^{\vec \mu}) \vec{\mathcal{H}} & =\sqrt{\mu} \vec j,
  \end{aligned}
\end{equation}
where $\vec{\mathcal{E}}=\sqrt{\epsilon} \vec E$ and $\vec{\mathcal{H}}=\sqrt{\mu}\vec H$. The new vector fields that appear by the right-hand multiplication operators $M^{\vec \epsilon}$ and $M^{\vec \mu}$ are given respectively by 
\begin{equation}
  \begin{aligned}
    \vec \epsilon & =\frac{\nabla \sqrt{\epsilon}}{\sqrt{\epsilon}}=\frac{\nabla \epsilon}{2 \epsilon}, \\
    \vec \mu & =\frac{\nabla \sqrt{\mu}}{\sqrt{\mu}}=\frac{\nabla \mu}{2 \mu}.
  \end{aligned}
\end{equation}
Obviously, \eqref{eq:Maxwell-1} and \eqref{eq:Maxwell-2} are equivalents.

It has been noticed \cite{Krav2003} that scalar fundamental solutions of the Schr\"{o}dinger operator with potential $\Delta \varphi/\varphi=c^2$ (where $c$ is a constant) generate purely vector fundamental solutions of the operator $D+M^{\vec \alpha}$, where $\vec \alpha=\nabla \varphi/\varphi$. Unfortunately, we cannot used this procedure to generate a fundamental solution in the present case. The advantage to know a fundamental solution of $D+M^{\vec \alpha}$ is that we could adapt the solution method presented in Section \ref{sec:The operator D+alpha}. Instead of that, we will apply Theorem \ref{th:solution-D+M} in order to give an explicit solution of the time-independent Maxwell system in inhomogeneous media \eqref{eq:Maxwell-1}.

\begin{theorem}
  Let $\Omega$ be a star-shaped domain. Let $\mu, \epsilon \in W^{1,\infty}(\Omega)$ be non vanishing scalar functions and $\rho,\vec j\in L^p(\Omega)$. Then a weak general solution of \eqref{eq:Maxwell-2} is given by
  \begin{equation}
    \label{eq:solution-Maxwell}
    \begin{aligned}
      \vec E & =-\nabla h_1+\frac{\vec u_1}{\sqrt{\epsilon}}, \\
      \vec H & =R_{\Omega}[\vec j]-\nabla h_2+\frac{\vec u_2}{\sqrt{\mu}},
    \end{aligned}
  \end{equation}
  where $\vec u_1$ and $\vec u_2$ are arbitrary vector fields in the kernel of $D+M^{\vec \epsilon}$ and $D+M^{\vec \mu}$, respectively. Moreover, $h_1$ and $h_2$ are respectively solutions of the conductivity equations
  \begin{equation}
    \begin{aligned}
      \div(\epsilon \nabla h_1) & =-\rho, \\
      \div(\mu \nabla h_2) & =\nabla \mu \cdot R_{\Omega}[\vec j].
    \end{aligned}
  \end{equation}
\end{theorem}

\begin{proof}
  By \eqref{eq:Maxwell-1}, it readily follows that $\div \vec j=0$ in the time-independent case. We will verify that the right-hand sides of the equations in \eqref{eq:Maxwell-2} satisfy the hypotheses of Theorem \ref{th:solution-D+M}. The fact that $\vec \epsilon $ and $\vec \mu$ are irrotational vector fields is straightforward, and the compatibility condition $\div (\sqrt{\mu}\vec j)=\vec \mu \cdot \sqrt{\mu}\vec j$ holds. Applying Theorem \ref{th:solution-D+M}, we have 
  \begin{equation}
    \begin{aligned}
      \vec{\mathcal{E}} & =-\sqrt{\epsilon} \nabla h_1+\vec u_1, \\
      \vec{\mathcal{H}} & =\sqrt{\mu} R_{\Omega}[\vec j]-\sqrt{\epsilon} \nabla h_2+\vec u_2,
    \end{aligned}
  \end{equation}
  where $h_1$ and $h_2$ are solutions of the conductivity equations
  \begin{equation}
    \begin{aligned}
      \div(\epsilon \nabla h_1) & =-\rho, \\
      \div(\mu \nabla h_2) & =\nabla \mu \cdot R_{\Omega}[\vec j].
    \end{aligned}
  \end{equation}
  Without loss of generality, suppose that $h_i|_{\partial\Omega}=0$ has zero trace for $i=1,2$. By the non-uniqueness of the solutions established in Theorem \ref{th:solution-D+M}, let $\vec u_i$ be such that $(D+M^{\epsilon})\vec u_1=0$ and $(D+M^{\mu})\vec u_2=0$, respectively.
Finally, the last expression comes from the fact that $\vec{\mathcal{E}}=\sqrt{\epsilon} \vec E$ and $\vec{\mathcal{H}}=\sqrt{\mu}\vec H$.
\end{proof}




\begin{thebibliography}{99}

\bibitem{ABDG1998} C.~Amrouche, C~Bernardi, M.~Dauge,
  V.~Girault. ``Vector Potentials in Three-Dimensional Nonsmooth
  Domains.'' {\em Math. Meth. Appl. Sci.,} \textbf{21} (1998)
  823--864.


\bibitem{Berselli2019} L.~C.~Berselli, P.~Longo, ``Classical solutions for the system $\curl v=g$, with vanishing Dirichlet boundary conditions'' {\em Discrete and Continuous Dynamical Systems}, 12 (2): (2019) 215--229.

\bibitem{Borchers1990}
W.~Borchers and H.~Sohr ``On the equations $\text{rot} v = g$ and $\div u = f$ with zero boundary conditions'', {\em Hokkaido Math. J.}, 19 (1990), 67--87.

\bibitem{Brezis2007} J.~Bourgain and H.~Brezis, ``New estimates for elliptic equations and Hodge type systems'', {\em J. Eur. Math. Soc. (JEMS)}, 9 (2007), 277--315.

\bibitem{Chandrasekhar1957}
S.~Chandrasekhar, P.~Kendall, ``On force-free magnetic fields''. Astrophysical Journal 126, 457--460 (1957).

\bibitem{DahKen1987} B.~Dahlberg, C.~Kenig, ``Hardy spaces and the Neumann problem in $L^p$ for Laplace's equation in Lipschitz
  domains.'' {\em Ann. of Math.} \textbf{125} (1987) 437--465.
	
\bibitem{Dautray1985} R.~Dautray, J.-L.~Lions, {\em Mathematical
    Analysis and Numerical Methods for Science and Technology}.  Vol.\
  3, Springer-Verlag, New York, Berlin (1985).
		
\bibitem{DelPor2017} B.~B.~Delgado, R.~M.~Porter,  ``General
solution of the inhomogeneous div-curl system and consequences,'' {\em Advances in Applied Clifford Algebras,} Vol. 27, Issue 4, 3015--3037 (2017) DOI 10.1007/s00006-017-0805-z.

\bibitem{DelPor2018} B.~B.~Delgado, R.~M.~Porter, ``Hilbert transform for the three-dimensional Vekua equation,'' {\em Complex Variables and Elliptic Equations} Vol. 64, No. 11, 1797--1824,  (2018) DOI 10.1080/17476933.2018.1555246. 

\bibitem{Thesis2018} B.~B.~Delgado. ``Quaternionic Vekua Analysis in Domains in $\R^3$ with Application to Electromagnetic Systems of Equations,'' {\em PhD Thesis} (2018). 

\bibitem {DelKrav2019}
B.~B.~Delgado, V.~V.~Kravchenko, \textquotedblleft A
right inverse operator for $\mathit{curl}+\lambda$ and
applications,\textquotedblright\emph{Advances in Applied Clifford Algebras, 29:40} (2019).
\bibitem{DelMac2021}
B.~B.~Delgado,J.~E.~Mac\'{\i}as-D\'{\i}az, ``An Exterior
Neumann Boundary-Value Problem for the Div-Curl System and Applications,'' \emph{Mathematics, 9(14): 1609} (2021).

\bibitem{FaJoLe1977} E.~Fabes, M.~Jodeit~Jr., and J.~Lewis, ``Double layer potentials for domains with corners and edges.''  {\em Indiana Univ.\ Math.\ J.}\ \textbf{26} (1977) 95--114.

\bibitem{Girault1986} V.~Girault and P.~A.~Raviart, {\em Finite Element Methods for the Navier-Stokes Equations}.
  Springer-Verlag, Berlin (1986).
	
 
\bibitem{Griesinger1990} R.~Griesinger. ``The boundary value problem \text{rot} $u = f$, $u$ vanishing at the boundary and the related decompositions of $L^q$ and $H_0^{1,q}$: existence''. {\em Ann. Univ. Ferrara Sez. VII (N.S.)}, 36 (1990), 15--43.


\bibitem{GuSpr1997} K.~G\"{u}rlebeck, W.~Spr\"{o}\ss{}ig.  {\em
    Quaternionic and Clifford Calculus for Physicists and Engineers}. Chichester: John Wiley \& Sons (1997).
	
\bibitem{GuHaSpr2008} K.~G\"{u}rlebeck, K.~Habetha,
W.~Spr\"{o}\ss{}ig.  {\em Holomorphic Functions in the Plane and
n-dimensional Space}.  Birkh\"{a}user, Basel (2008).

\bibitem{GuHaSpr2016} K.~G\"{u}rlebeck, K.~Habetha,
  W.~Spr\"{o}\ss{}ig, {\em Application of Holomorphic Functions in Two
    and Higher Dimensions}. Birkh\"{a}user, Basel (2016).

\bibitem{Helmholtz1873} H.~Helmholtz, ``Ueber die Theorie der Elektrodynamik.'' {\em Zweite Abhandlung. Kritisches, J.
Reine Angew. Math.,} 75 (1873), 35--66.

\bibitem{Hiptmair2012}
R.~Hiptmair, P.~R.~Kotiuga, S.~Tordeux. ``Self-adjoint curl operators''. {\em Annali di Matematica}. \textbf{191} 431--457 (2012).

\bibitem{Isakov1998} V.~Isakov,  {\em Inverse problems for partial differential equations}.  Springer-Verlag, Berlin (1998).

\bibitem{JeKe1981} D.~S.~Jerison and C.~E.~Kenig,  ``The Dirichlet problem in nonsmooth domains.''  {\em Ann. of Math.} \textbf{113} (1981) 367--382.

\bibitem{Jette1970}
A.~Jette. ``Force-free magnetic fields in resistive magnetohydrostatics''. J. Math. Anal. Appl. 29, 109--
122 (1970).
	
\bibitem{Korn1968} G.~A.~Korn, T.~M.~Korn, {\em Mathematical Handbook for Scientists and Engineers}.  Dover Publications, Inc, Mineola (1968).		

\bibitem{Krav2003} V.~V.~Kravchenko.  {\em Applied Quaternionic Analysis}.  Heldermann Verlag: Lemgo (2003).

\bibitem{Krav2002} V.~V.~Kravchenko. ``Quaternionic reformulation of Maxwell's equations for inhomogeneous media and new solutions.'' {\em Zeitschrift F\"{u}r Analysis und ihre Anwendungen.} 21 (2002) 21--26.


\bibitem{kravchenko2021complete}
  V.~V.~Kravchenko, P.~E.~Moreira and M.~R.~Porter,
  ``Complete Systems of Beltrami Fields Using Complex Quaternions and Transmutation Theory'', \em{Advances in Applied Clifford Algebras}, 31:3, (2021) 1--21.
  
\bibitem{Kub2012} C.~S.~Kubrusly, {\em Spectral Theory of Operators on Hilbert Spaces}. Birkh\"{a}user, New York (2012).

\bibitem{Mclean2000} V.~McLean, {\em Strongly Elliptic Systems and Boundary Integral Operators}.Cambridge University Press, 1st edition (2000). 

\bibitem{Mikhailov1978} V.~P.~Mikhailov,  {\em Partial differential equations}.  Mir Publishers, Moscow (1978).

\bibitem{Mitrea1994} M.~Mitrea.  {\em Clifford Wavelets, Singular integrals, and Hardy spaces}.  Lectures Notes in Mathematics, vol. 1575, Springer-Verlag, Berlin (1994).

\bibitem{Simader1972} C.~G.~Simader, {\em On Dirichlet’s boundary value problem}, Lecture Notes in Mathematics
268, Springer 1972.

\bibitem{Sprossig1993} W.~Spr\"{o}ssig, ``On the treatment of Non-linear Boundary Value Problems of a Disturbed Dirac Equation by Hypercomplex Methods.''  {\em Complex Variables,} Vol. 23 (1993) 123--130.

\bibitem{Sprossig2001} W.~Spr\"{o}ssig, ``On generalized Vekua type problems.''  {\em Advances in Applied Clifford algebras.} \textbf{11} (2001) 77--92.

\bibitem{Temam1979}
R.~Temam. {\em Navier-Stokes Equations. Theory and Numerical Analysis}, North-Holland, Amsterdam, (1979).

\bibitem{Ver1984} G.~Verchota, ``Layer Potentials and Regularity for
the Dirichlet Problem for Laplace's Equation in Lipschitz Domains.''
{\em J.\ Funct.\ Anal.}\ \textbf{59} (1984) 572--611.

\bibitem{Wahl1990} W.~von~Wahl, ``On necessary and sufficient conditions for the solvability of the equations
$\text{rot } u = \gamma$ and $\text{div } u = \epsilon$ with u vanishing on the boundary'', in The Navier-Stokes equations (Oberwolfach, 1988), vol. 1431 of Lecture Notes in Math., Springer, Berlin, (1990), 152--157.

\bibitem{Weber1980} A.~Ch.~Weber, ``A Local Compactness Theorem for Maxwell's Equations.''  {\em Math. Meth. in the Appl. Sci.}
\textbf{2} (1980) 12-25.

\bibitem{Yoshida1990}
Z.~Yoshida, Y.~Giga, ``Remarks on spectra of operator rot''. {\em Math. Z.} 204, 235--245 (1990).

\end{thebibliography}
\end{document}